\documentclass[12pt]{iopart}
\usepackage{fullpage}

\usepackage{graphicx}
\usepackage{fixltx2e} 
\usepackage{epsfig} 
\usepackage{epstopdf}
\usepackage{textgreek} 
\usepackage[usenames,dvipsnames]{color}
\usepackage[usenames,dvipsnames,svgnames,table]{xcolor}
\usepackage[normalem]{ulem} 
\DeclareMathAlphabet{\mathpzc}{OT1}{pzc}{m}{it}

\usepackage{hyperref}
\hypersetup{colorlinks=true,linkcolor=blue,anchorcolor=blue,citecolor=blue,filecolor=blue,urlcolor=blue,bookmarksnumbered=true,pdfview=FitB}

\definecolor{gray}{gray}{0.8}
\definecolor{darkgray}{gray}{0.6}

           \newcommand{\Hb}{{\mathbf H}}

           \newcommand{\Mb}{{\mathbf M}}

\newcommand{\rb}{{\mathbf r}}

\newcommand{\vb}{{\mathbf v}}

\usepackage{dsfont}

\DeclareMathAlphabet{\mathpzc}{OT1}{pzc}{m}{it}

\newcommand{\TR    }{{\top}}

\newcommand{\Ca}{\ensuremath{\mbox{C}_\alpha}}

\newcommand{\NMA  }{{\mbox{\scriptsize NMA}}}

\newcommand{\spr   }{{\mbox{\scriptsize spr}}}
\newcommand{\frc   }{{\mbox{\scriptsize frc}}}

\begin{document}

\date{}
\title[Universality of Vibrational Spectra of Globular Proteins]
  {Universality of Vibrational Spectra of Globular Proteins}

\author{Hyuntae Na$^{1}$, Guang Song$^{1,2,3}$, and Daniel ben-Avraham$^{4}$}

\address{$^1$ Department of Computer Science,} 
\address{$^2$ Graduate Program of Bioinformatics and Computational Biology,} 
\address{$^3$ L. H. Baker Center for Bioinformatics and Biological Statistics, \\ Iowa State University, 226 Atanasoff Hall, Ames, Iowa 50011, USA}
\address{$^4$ Department of Physics, Clarkson University,  Science Center 265, Potsdam, NY 13699, USA\\
~\\
Correspondence: Guang Song, Tel: 515-294-1696, Fax: 515-294-0258
}
\eads{\mailto{gsong@iastate.edu}}

\begin{abstract}
It is shown that the density of modes of the vibrational spectrum of globular proteins is universal, i.e., regardless of the protein in question, it closely follows one universal curve. 
The present study, including 135 proteins analyzed with a full atomic empirical potential (CHARMM22) and using the full complement of all atoms Cartesian degrees of freedom, goes far beyond previous claims of universality, {confirming} that universality holds even in the {frequency range that is well above $100\,{\rm cm}^{-1}$}  ($300$ -- $4000\,{\rm cm}^{-1}$), where peaks and turns in the density of states are faithfully reproduced from one protein to the next.  
{We also characterize fluctuations of the spectral density from the average, paving the way to a meaningful discussion of rare, unusual spectra and the structural reasons for the deviations in such ``outlier" proteins. 
{Since} the method used for the derivation of the vibrational modes (potential energy formulation, set of degrees of freedom employed, etc.) has a dramatic effect on the spectral density, another significant implication of our findings is that the universality can provide an exquisite tool for assessing and improving the quality of potential functions and the quality of various models  used for NMA computations.
Finally, we show that the input configuration too affects the density of modes, thus emphasizing the importance of simplified potential energy formulations that are minimized at the outset.
In summary,} our findings call for a serious two-way dialogue between theory and experiment: Experimental spectra of proteins could now guide the fine tuning of theoretical empirical potentials, and the various features and peaks observed in theoretical studies -- being universal, and hence now rising in importance -- would hopefully spur experimental confirmation.
\end{abstract}

\vspace{2pc}
\noindent{\it Keywords}: Universality; Vibrational Spectrum; Normal Modes; {Globular Proteins; Folds}; Density of Modes; Force Field; Protein Dynamics; NMA

\maketitle


\section{Introduction}
\label{sect:introduction}

The atomic structures of thousands of proteins have been elucidated and display recurring patterns of folding,
such as the common globin and Greek key folds, and the $\beta$-barrel folds.  These repeating structural
motifs obtain distinct flexibility signatures.  Identifying these intrinsic deformability characteristics
is required to ascertain and better understand protein functionality.  Historically, the characterization of any object's internal
deformabilities under small perturbations has been achieved by a normal mode analysis of its internal
degrees of freedom.  While a normal mode analysis is a well-defined and straightforward computation,
the identification of a suitable set of internal degrees of freedom and an appropriate
 potential energy formulation to quantify the effects of deformations, remains more of an art.  
 Here we examine the  spectrum of vibrations obtained for a large number of proteins, using several of the more traditional approaches for normal mode analysis.  {W}ithin a given approach, the density of the spectrum of vibrations is universal, despite the many significant differences among individual proteins.

Normal modes of proteins have been studied since the early 1980's~\cite{Tasumi82,Go83,Brooks83,Levitt83}.   A normal mode calculation requires as input an empirical potential function for the various forces between the protein's atoms:  the more detailed the atomic potential function, the more reliable the results, but on expense of more cumbersome computations.  Starting with the seminal work of Tirion~\cite{Tirion96}, various simpler alternatives to a detailed potential have been explored~\cite{Bahar97,Bahar97b,Hinsen98,Hinsen99,Atilgan01,Tama01,
Li02,Ma04,VanWynsberghe05,Zheng05,Yang09,Lin10,Na14b,Na15a,Tirion15}.
In addition to simplifications in the potential energy formulation, reduced sets of
internal degrees of freedom (dofs) have been explored.  Two common choices include the restricted set of dihedral angles degrees of freedom, or {\em torsional} dofs, for short (the rationale being that changes in bond lengths and angles require far larger energy investment than dihedral or torsional changes)~\cite{Go83,Levitt83,Tirion93} and Cartesian dofs for the reduced set of only the \Ca\,atoms~\cite{Bahar97,Bahar97b}.

Early normal mode analyses examined the density of the modes by frequency range, $g(\omega)$, and deliberated the meaning of the various features in the curves found for each protein.  However, it was soon found out that,
when properly normalized, the $g(\omega)$ of different proteins seem to collapse onto one universal curve,
characteristic of globular proteins in general~\cite{Tirion93,dba93}.  This initial finding was based on merely 5
proteins, and on a normal mode analysis with only torsional dofs.  {{Most} later studies of the distribution of normal modes
did little to confirm the universality of $g(\omega)$, as they focused on properties of the spectrum only at the
low frequency range (up to $\sim 20\,{\rm cm}^{-1}$) and tended to rely on simplified potential functions.
{An exception is the recent analyses of Hinsen et al.,~\cite{Hinsen00-ms,Hinsen05} of crambin, lysozyme, and myoglobin, using the AMBER potential, that suggested that the universality of the density of the modes extends to all frequencies.}

In this paper, we re-examine the hypothesis that $g(\omega)$ is universal.  Advances in computer technology
in recent years 
allow us to consider $135$ globular protein structures
whose resolutions are better than 2.5~\AA~and whose sequence identity is less than 30\%, 
and obtain spectra of normal modes with a detailed atomistic empirical potential  
and the full complement of Cartesian degrees of freedom. (Some of our results are presented for torsional dofs only and/or for simplified potentials.)  This wealth of information lets us do much more than simply confirm
the putative universality of $g(\omega)$:  (1)~Our main result is that the density of the spectrum of vibrations, $g(\omega)$, is universal also for the full complement of Cartesian dofs, down to the seemingly idiosyncratic peaks and details in the high-frequency range. 
This is a big surprise: in the low-frequency range universality is expected on the grounds that slow modes involve long-wavelength coherent motion of large domains of a protein, and therefore the many interactions involved (at the surfaces between domains) average out in the same fashion, regardless of details.  In contrast, high-frequency oscillations involve small coherence lengths
and motions of {\em small} groups of atoms relative to one another, so here universality is unexpected.   
(2)~Our large data set allows us to characterize not only a reliable average for $g(\omega)$, but also the typical fluctuations from that average.  Specific features in the $g(\omega)$ of a protein are unusual only in comparison to these fluctuations, so the old notion of identifying and discussing the meaning of salient features of $g(\omega)$ of a protein finally becomes possible. For example, our data allows us to identify subtle, yet meaningful differences in the spectra of proteins of different folds. Some of these observations are echoed in experimental findings. (3)~The universal curve for $g(\omega)$ depends on the specific empirical potential one uses, its parameters, etc., whether the potential is detailed or simplified, as well as on the set of degrees of freedom (e.g., Cartesian or torsional).  We show that the comparison of the $g(\omega)$'s arising in each case is a very sensitive way to assess the accuracy and success of the various approximations and approaches.  (4)~Working with an atomistic detailed potential, the first step in an NMA involves minimizing the potential function, thereby altering the input PDB 
structure.  In other simplified approaches,
one posits a potential that is minimized at the given configuration (PDB, or other) at the outset. We show that energy-minimized starting configurations obtain significantly different spectra $g(\omega)$ than the original PDB starting configurations, and we discuss the implications of this finding.

The remainder of this paper is organized as follows.  In Section~\ref{methods:sec} we describe our protein dataset and briefly review the theoretical technique of normal modes analysis and the various approaches
(full and simplified potential functions, choices of dofs, etc.) considered in this work.  Our results are presented
and analyzed in Section~\ref{results:sec}.  Final conclusions and promising open problems are discussed
in Section~\ref{conclusion:sec}.


\section{Materials and Methods}
\label{methods:sec}
\label{hypref:methods:sec}
\subsection{The Protein Dataset}

The protein dataset used in this study is the same as the one used in a previous work by Na and Song~\cite{Na15c}. The dataset includes 135 proteins resolved to better than {{2.5\,\AA}} and following minimization none of the proteins undergoes more than a {{6.0\,\AA}} RMSD change.  The proteins are quite evenly divided between different fold classes, including 42  all-$\alpha$ proteins, 37  all-$\beta$ proteins, and 56  $\alpha/\beta$-proteins. Their sizes range from 61 residues (pdb-ids: 1I2T, 1I0M, 2J5Y, 3MP9) to 149 residues (pdb-ids: 1GU1, 2Y9F, 3AXC); the distribution of the proteins by size is illustrated in figure~\ref{fig:sizes}{(A)}. Only small to medium sized proteins are used here due to the large computational cost of running NMA.
 The protein structures are energetically minimized using the Tinker program~\cite{Tinker} with the CHARMM22 force field~\cite{MacKerell98}. {The amount of structure deviations due to energy minimization is given in figure~\ref{fig:sizes}(B). No cutoff distance is specified in the process. As a result, the program does not taper the electrostatic or the van der Waals potential with any smoothing function but considers all pair-wise non-bonded interactions.} 
The minimized structures and the original PDB structures are  available at
\url{http://www.cs.iastate.edu/~gsong/CSB/NMAdb/135.html}.

\subsection{Normal Modes Analysis}

Normal modes analysis (NMA) was first applied to proteins in the early 80's.~\cite{Tasumi82,Go83,Brooks83,Levitt83}. Conventional NMA proceeds from a detailed atomic potential function, $V$, for the interactions within the protein system.  Generally, the input structure (mostly a PDB structure) is not at an energy minimum according to $V$. As required by NMA, 
the potential function has to first be minimized --- a computationally expensive operation that also distorts the starting configuration by as many as several angstroms (RMSD).

Using the minimized structure, one constructs the Hessian matrix ${\bf H}$, which is the second derivative of the potential energy with respect to the protein's degrees of freedom $\{q_i\}_{i=1}^N$;
\begin{equation}
{\rm H}_{ij}=\frac{\partial^2V}{\partial q_i\partial q_j}\,,
\end{equation}
as well as the mass, or inertia matrix ${\bf M}$;
\begin{equation}
{\rm M}_{ij}=\sum_k m_k\frac{\partial{\bf r}_k}{\partial q_i}\cdot\frac{\partial{\bf r}_k}{\partial q_j}\,,
\end{equation}
where the sum runs over all atoms $k$ of the protein, and $m_k$ and ${\bf r}_k$ are the $k$-atom's  mass and location, respectively.  One then solves the generalized eigenvalue problem:
\begin{equation}
\Hb \vb_i = \lambda_i \Mb \vb_i\,.
\end{equation}
Here $\vb_i$ is the $i$-th vibrational eigenmode, and the eigenvalue $\lambda_i=\omega_i^2$ encodes its (angular) frequency;
$\vb_i(t)=\vb_i(0)\cos(\omega_i t)$.   For comparison with experimental work, it is customary to express $\omega_i$ in terms of the corresponding inverse wavelength of electromagnetic radiation (measured in ${\rm cm}^{-1}$). This is achieved by dividing its value (in radians/sec) by $2\pi c$, where $c$ is the speed of light, $c=2.997925\times10^{10}\,{\rm cm}/{\rm sec}$.

Finally, we compute the density of vibrational modes $g(\omega)$, the focus of this work, in the following way.  Subdivide the frequency range into bins of width $\Delta\omega$ and count the number of modes $n_j$ that have frequency $\omega$ within the $j$-th bin, i.e., modes with $\omega_j-\frac{1}{2}\Delta\omega<\omega<\omega_j+\frac{1}{2}\Delta\omega$.  Then $g(\omega_j)=n_j/(N\Delta\omega)$, where $N$ is the total number of dofs for the protein.  (Typically, $\Delta\omega=5$ or $10\,{\rm cm}^{-1}$.)  This procedure is done for each of the 135 proteins in our dataset and their $g(\omega)$'s are processed as needed (averaged, compared to one another, etc.)

There exist several choices for the $\{q_i\}_{i=1}^N$ degrees of freedom.  The simplest choice, conceptually,
is the full complement of Cartesian degrees of freedom; the $(x,y,z)$ coordinates for each of the atoms in the protein.  This yields a diagonal $\Mb$ matrix and minimization is conceptually simpler than with generalized dofs, but the number of dofs $N$ can become prohibitively large.  A common alternative in early work was using the quite smaller set of torsional and dihedral angle dofs; bond lengths and bond angles are frozen in this method (approximating the fact that these are much stiffer than the torsional and dihedral dofs).  A drawback of torsional dofs is that the $\Mb$ matrix is more complicated (though smaller), and the minimization algorithm is trickier.
In the present study our torsional modes are obtained by projecting a Cartesian Hessian onto the torsional space, as done in the Torsional Network Model~\cite{Mendez10}.

\subsection{Simplified Normal Mode Analyses}
\label{sect:ENM}
\label{hypref:sect:ENM}

Because conventional NMA is cumbersome to use, due to its complicated all-atom potential and energy minimization process, in 1996 Tirion proposed a simplified potential that required no minimization~\cite{Tirion96}.
Tirion's approach uses torsional dofs (freezing bond lengths and angles) and postulates a universal Hookean
potential between non-bonded atom pairs:
\begin{equation}
V=\sum_{\langle ij\rangle} \frac{1}{2}C(r_{ij}-r_{ij}^0)^2\,,
\end{equation}
where $r_{ij}$ and $r_{ij}^0$ are the current and the starting-configuration distance between atoms $i$ and $j$, respectively, and the sum runs over all $\langle i j\rangle$ non-bonded atom pairs that are sufficiently close to one another: $r_{ij}^0<r_{VdW}^i+r_{VdW}^j+r_c$ ($r_{VdW}^a$ is the Van der Waals radius of atom $a$ and $r_c$ is a cutoff distance,
typically a few angstroms).  The big advantage of the Tirion potential is that it requires no minimization, it is minimized at the outset at the starting configuration $\{\rb_i^0\}$.  Other potentials and approaches that require no minimization (known generally as {\em elastic network models}) have been developed since Tirion's seminal work.  We now review the main ingredients of the simplified approaches discussed in this paper.

{\bf ANM.} The ANM, or Anisotropic Network Model, was developed by Atilgan et al.,~\cite{Atilgan01} in 2001. It is mainly a coarse-grained version of the Tirion potential, with each residue represented only by its \Ca\,atom. It has been used also as an all-atom model, though to a much lesser extent.  Simplifying further still, ANM employs the easier to use Cartesian dofs.  However,  on obliterating the constraints of bond lengths and angles it washes out the distinction between bonded and non-bonded interactions and further loses in accuracy.
Because of its easy implementation ANM has been widely used in many normal mode-based studies and analyses.

{\bf sbNMA.}  In 2014, Na and Song~\cite{Na14a, Na14b} developed a new way for deriving simplified models for normal mode computations. They employed a top-down approach and derived several high-quality elastic network models (i.e., require no minimization) by gradually simplifying the conventional NMA.  The most accurate of these approaches is the spring-based NMA (sbNMA).  Structurally, sbNMA is the same as the conventional NMA and is an all-atom model. The interaction model of sbNMA, on the other hand, is different from the NMA force field from which it is derived.
While the Hessian in general consists of spring{-constant}-based  terms $\Hb^\spr$ {(or spring-based, for short)} { --- terms that are proportional to the spring constants ---} and  force/torque-based terms $\Hb^\frc$ {(terms that are proportional to the inter-atomic forces or torques)}~\cite{Na14a,Na14b}, {i.e.,} $\Hb^\NMA = \Hb^\spr + \Hb^\frc$, sbNMA keeps only the spring-based terms. The rationale was that the force/torque terms contribute significantly less to the overall dynamics than the spring-based terms~\cite{Na14b}. To ensure stability, regions where the spring constants become negative are excluded. 
For example,
electrostatic interactions (which were shown to contribute much less than van der Waals interactions~\cite{Na14b}), are not included, as attractive forces give rise to negative spring constants.
The spring-based NMA (or sbNMA) preserves much of the complexity of the original NMA, and the neglect of the force/torque terms has minimal impact.  As a result, sbNMA 
yields very high-quality vibrational modes and 
closely resembles NMA~\cite{Na14b}.

A very similar approach to sbNMA, dubbed ATMAN (for Atomic Torsional Mode Analysis), was developed independently by Tirion and ben-Avraham~\cite{Tirion15}.  ATMAN too keeps only spring-based terms, derived from a detailed atomic potential, and only wherever these are positive.  The main difference to sbNMA is that ATMAN
allows for ``stretching'' the range of positive spring constants, to compensate for the loss of range where the spring constants are negative.  This, however, adds tunable parameters.

{\bf ssNMA.} A further simplification beyond sbNMA is achieved by the simplified spring-based NMA, or ssNMA. It combines many of the different constants in sbNMA into one single parameter, thus requiring a much smaller set, of 17 parameters in total. For example, it uses a single bond-stretching spring constant for all bonded pairs of atoms, regardless of their types.  This of course results in some loss of accuracy, compared to sbNMA.  Below, we limit our study to NMA (with the CHARMM22 force field), the sbNMA and ssNMA derived from it, and ANM.  Note that only the original NMA requires minimization, while all of the simplified approaches can start from any given protein configuration.

\subsection{Computing the Contribution from Various Interaction Types}
\label{contributions:sec}
\label{hypref:contributions:sec}

The CHARMM22 potential energy function, which we use for NMA and sbNMA, consists of several types of terms: (a)~bond stretching, (b)~bond-angle bending, (c)~improper angle distortions, (d)~torsional and dihedral rotations, and (e)~non-bonded interactions, including Van der Waals and electrostatic forces. We group Urey-Bradley interactions along with the bond-angle bending terms.
One can compute the relative individual contribution from each type of interaction as follows. Since $V=V_{\rm bond}+V_{\rm angle}+\cdots+V_{\rm nonbonded}$, the Hessian decomposes into mutually exclusive matrices;
\begin{equation}
\label{Hs:eq}
\Hb=\Hb_{\rm bond}+\Hb_{\rm angle}+\Hb_{\rm improper}+\Hb_{\rm torsional}+\Hb_{\rm nonbonded}\,.
\end{equation}
Then, the relative contribution $c_{ij}$ of interaction type $j\in\{{\rm bond}, {\rm angle}, {\rm improper}, {\rm torsional}, {\rm nonbonded}\}$ to the $i$th mode $\vb_i$, is
\begin{equation}
c_{ij} = \vb_i^\TR \Hb_j \vb_i\,.
\end{equation}
Note that the $c_{ij}$ are properly normalized, $\sum_jc_{ij}=1$, because of equation~(\ref{Hs:eq}) and the fact that our eigenvectors are $\Hb$-normalized; $\vb_i^\TR\Hb\vb_i=1$. $c_{ij}$ is guaranteed to be greater or equal to 0 with approaches like sbNMA and ATMAN, where the various $\Hb_j$ are positive semi-definite.  Intuitively, $c_{ij}$ reveals the extent to which interaction type $j$ constrains the vibration along mode $i$.


\section{Results}
\label{results:sec}
\label{hypref:results:sec}

\subsection{Universality of the Density of Vibrational Modes}
\label{universality:subsec}
\label{hypref:universality:subsec}

\paragraph{Universality in the full complement of Cartesian dofs.}

Our main result is presented in figure~\ref{fig:universal}, which demonstrates that the density of vibrational modes for each of the proteins in our dataset is very nearly the same.  In other words, {\em the density of the spectrum of vibrations is universal}. To obtain this figure, we have conducted a full NMA on each of the proteins in the dataset, employing the CHARMM22 atomic potential function, and using all of the Cartesian dofs of each protein's atoms, and obtained their $g(\omega)$'s.   In the plot, we show the average of $g(\omega)$ over all 135 proteins (black curve); fluctuations from the average are indicated by colored bands that include proteins within different percentile ranges: 25--75 percentile (orange), 5--95 percentile (red), 0--100 percentile (gray).  An animation that displays the vibrational spectra of the proteins one by one and illustrates how they all share a common spectrum pattern is given in the Supporting Information.

Surprisingly, even accounting for extreme fluctuations (the 0--100 percentile includes {\em all} of the proteins in the dataset), the various main features of $g(\omega)$ --- seemingly idiosyncratic turns and peaks --- are faithfully reproduced throughout the whole frequency range.
These peaks  must thus correspond to some physical characteristics of the structure of globular proteins in general, and to physical interactions within them that are {independent} of the details of each individual protein structure.

The universality of $g(\omega)$ would seem to exclude any possibility of gleaning particular knowledge of a protein from its specific vibrational spectrum.  This is not necessarily so: If the density of modes of a protein deviates {\em significantly} from the average $g(\omega)$, perhaps the deviations can tell us something about the structure of the specific protein in question.  Whether a deviation is significant, could  be decided from the fluctuation bands in figure~\ref{fig:universal}.  There could be, however, a different cause for deviations, besides anomalous structure: Relative fluctuations of a random variable tend to decrease inversely proportional to the square root of the system's size (the protein's size), so smaller proteins would exhibit larger fluctuations, even if their structure
is not anomalous.  With that in mind, we have examined the proteins that fall outside of the 5--95 percentile, but found no correlation with size: The distribution of protein sizes in that range is very similar to that of the dataset at large (figure~\ref{fig:sizes}(A)).   Based on these preliminary results, we conclude that rare fluctuations are more likely due to structural anomalies.  A more detailed study of the dependence (or independence) of fluctuations on size,
and of rare outlier proteins is left for future work.

We stress that ours is the first study of universality of $g(\omega)$ that employs the full complement of Cartesian degrees of freedom.  Universality was first discovered by studying the spectrum from the restricted set of torsional dofs, in the 0--300$\,{\rm cm}^{-1}$ frequency range~\cite{Tirion93,dba93}.  For low frequencies, universality is expected because slow modes involve motion of large domains of a protein, and  the many interactions at the surfaces between domains average out uniformly, by the law of large numbers, regardless of details.  High-frequency oscillations, on the other hand, involve small coherence lengths and only a few atoms, so the law of large numbers cannot be as easily  invoked. We were therefore unprepared for the striking results of figure~\ref{fig:universal}. 
{A possible explanation is that, because the high-frequency modes represent the oscillations of only a few atoms relative to the rest of the protein, that has a much bigger mass, the size and structural details of the rest of the protein is mostly irrelevant and consequently the characteristics of the oscillations depend only on the structural composition of the few oscillating atoms and are largely protein independent.} We also note that the scope of the present study, of 135 proteins, far surpasses the scope of previous works on the subject, proving universality beyond all doubt, at least from a theoretical point of view.
Experimental study on four proteins that have different mixtures of secondary-structures showed that their vibrational spectra have ``a common appearance''~\cite{Giraud03}.
This experimental result in the low-frequency range thus seems to confirm universality as well~\cite{Giraud03}
(see also Section~\ref{subsec:potentials}).

\paragraph{Cartesian vs. torsional degrees of freedom.}
\label{torsional:subsec}
\label{hypref:torsional:subsec}

We now examine the density of vibrations for the restricted set of torsional dofs.  Recall that changes in torsional angles are a lot easier to effect than changes in bond lengths and angles, so most of the protein's motion under thermal excitation could be accounted by torsional changes alone.  Moreover, because the reduced set of torsional angles is much smaller than the full complement of Cartesian dofs, it allows analysis of much larger proteins and protein systems, a fact that accounts for much of their popularity.

In figure~\ref{fig:Cartesian-torsional} we compare the vibrational modes in Cartesian dofs (dashed black lines) to those in torsional dofs (solid black lines) for four randomly selected proteins in our dataset: 3NBC~(a), 3RHB~(b), 2QCP~(c), 3MP9~(d).
Since the total number of dofs is different for Cartesian and torsional coordinates, we show the actual count of modes in each $5\,{\rm cm}^{-1}$-bin, instead of the usual density of modes.  In this way one can see that the additional bond-length and angle bending dofs included in the Cartesian complement add vibrational modes in 
each bin.  Nevertheless, there is a clear relationship between the Cartesian and torsional curves:  Both the {\em main peak}, at about $80\,{\rm cm}^{-1}$, and the {\em secondary peak}, at about $300\,{\rm cm}^{-1}$, are in nice agreement
in the two representations (albeit with more modes present in the Cartesian dofs). As the frequency increases, torsional modes die out and a comparison becomes irrelevant.  The first two peaks are particularly important, as they encompass the low frequencies that account for most of the protein's thermal motions (B-factors, etc.).

Same as for the Cartesian case, the density of torsional modes for the proteins in the dataset  clusters around a single universal curve (the solid red curve in figure~\ref{fig:Cartesian-torsional}, which is the averaged density of modes over all the proteins in the dataset).
Knowing that $g(\omega)$ is universal in either choice of dofs, Cartesian or torsional, and that the two representations agree on the first two low-frequency peaks, allows one to use whichever set of dofs is convenient for the question at hand.  Indeed, in what follows, we shift back and forth between these two choices.

\paragraph{The origin of the peaks.}
\label{peaks:subsec}
\label{hypref:peaks:subsec}

We now address the question of what gives rise to the various detailed features in the spectrum density $g(\omega)$.  For each particular eigen-frequency, $\omega_i$, we compute, using the sbNMA model, the relative contribution $c_{ij}$ ($j=a,b,\dots,e$) of each of five interaction types: (a)~bond stretching, (b)~{bond-angle} bending, (c)~improper angle distortions, (d)~torsional and dihedral rotations, and (e)~non-bonded interactions, as explained in Section~\ref{contributions:sec}.
The contribution of interactions of type $j$ to the spectrum density is then given by $c_{ij}g(\omega_i)$.  The relative contributions of the five interaction types are shown in figure~\ref{fig:peaks}.

Of the five types of interaction only {{(d) and (e)}} are accessible with torsional dofs, since torsional and dihedral angle changes can affect neither bond lengths nor bond angles.  Thus one expects those two types of interaction to account for all of the spectrum density observed with the restricted set of torsional dofs.  Indeed, {{the dark gray- and light gray-shaded}} regions in figure~\ref{fig:peaks}, corresponding to the two types {{(d) and (e) respectively}}, bear a remarkable resemblance to the torsional spectra of figure~\ref{fig:Cartesian-torsional}.

Both the main and secondary peaks originate largely from torsional motions. The first peak is mainly due to torsional motions along the backbone, the $\phi$ or $\psi$ rotations, since 
{they involve the motions of larger masses}
and hence tend to oscillate at lower frequencies, while the second peak is probably more influenced by the torsional motions of the side-chain rotamers, which is of higher frequency since the mass of the rotating component is smaller. {This is in agreement with a previous study that suggested that the low frequency modes are contributed mainly by rigid body motions of the entire residues and side chain rotations~\cite{Hinsen00-ms}.}

As the frequency increases beyond about $500\,{\rm cm}^{-1}$ the bond stretching and bond angle interactions account for the lion's share of the spectrum density.  The peaks at about 1,000 -- 1,700$\,{\rm cm}^{-1}$ are largely dominated by angle bending interactions, while those in the 3,000 -- 4,000$\,{\rm cm}^{-1}$ region arise mostly from bond stretching.  Indeed, experimental results~\cite{MacKerell98,Palmo03} involving gas-phase infrared and Raman spectroscopy, though
limited to small molecules such as N-methylacetamide (CH\textsubscript{3}-CO-NH-CH\textsubscript{3}) or alanine dipeptide, confirm the vibrations of bonds and angles at frequencies similar to the peaks we find, and have been even used for adjusting force-field parameters.  The additional (universal) information provided by the spectrum density $g(\omega)$ {of globular proteins}, combined
with such experiments, could help better fine-tune the existing empirical potential energy functions (see, also, Section~\ref{subsec:potentials}).

One can confirm the origin of the main and secondary peaks in yet another, perhaps more direct way.  In figure~\ref{fig:torsional} we plot the torsional dofs spectrum density of sbNMA  under three different conditions: (i)~The original sbNMA, in gray (NMA is shown in black, for reference), (ii)~without the torsional and dihedral interaction terms (blue), and (iii)~ with those terms included, but without the non-bonded interaction terms (red).  The elimination of the torsional interactions results indeed in the obliteration of the secondary peak, and the elimination of non-bonded interactions results in a big distortion of the main peak, in agreement with the conclusions of the foregoing analysis.  Notice also that the correspondence between main peak and non-bonded interactions, and secondary peak and torsional interactions is not quite perfect: Elimination of torsional terms has some effect on the shape of the main peak as well, and elimination of  the non-bonded interactions seems to have quite a dramatic effect not only on the main peak but also on the secondary peak.  This agrees nicely with  figure~\ref{fig:peaks}),
where the shading indicates that both non-bonded interactions and torsional terms contribute to the main and secondary peaks, though to a different extent (the relative torsional contribution is larger for the secondary peak).

\subsection{Vibrational Spectra for Different Protein Folds}
\label{subsec:folds}
\label{hypref:subsec:folds}

Are there different vibrational spectra for proteins belonging to different classes of fold?  It has long been known~\cite{Levitt85} that secondary elements such as $\alpha$-helices, $\beta$-sheets, and turns exhibit different typical vibrational frequencies in the range below $\sim100\,{\rm cm}^{-1}$. These differences should in principle show in the vibrational spectra of proteins belonging to different fold classes, though the typical fluctuations from one protein to the next (figure~\ref{fig:universal}) would seem to impose a formidable obstacle to observing this phenomenon.  Indeed, the spectra of any two proteins in our data set belonging to a different fold, an all-$\alpha$ protein and an all-$\beta$ protein, say,  are clearly within the typical fluctuations range.  This result is in accord with experimental
findings of Giraud et al.,~\cite{Giraud03} who failed to observe any significant differences between the spectra of $\alpha$-rich and
$\beta$-rich proteins in their studies with ultrafast
OHD-RIKES spectroscopy. With some care,
however, one can tease out small but significant differences from our theoretical simulations.

Since the differences, if any,  are expected in the low-frequency range, we carry our analysis in only torsional dofs, which capture the features of $g(\omega)$ in this range most cleanly.  We have used the CATH catalog of protein structure {classification~\cite{Sillitoe15}} for identifying the fold type of most of the proteins in our dataset;  the remainder of the proteins were sorted out by manual inspection.  We thus identified 42 alpha-proteins (27 by CATH and 15 manually), 37 beta-proteins (27 by CATH and 10 manually), and 56 $\alpha/\beta$-proteins (38 by CATH and 18 manually).  We then averaged the torsional vibrational frequency distributions $g(\omega)$ within each group.
The three resulting curves are plotted in different colors in figure~\ref{fig:folds}a (all-$\alpha$ in red, all-$\beta$ in blue, and $\alpha/\beta$ in gray).  Although the three curves are very similar to one another, systematic deviations can be clearly seen upon closer inspection: The main peak shifts progressively to the left, from all-$\alpha$ to $\alpha/\beta$- to all-$\beta$ proteins, and the opposite trend occurs at the farther slope of the secondary peak.  That these shifts are systematic can be most clearly seen by randomly reassigning the proteins in the dataset to three groups of corresponding sizes, and recomputing the averages in the random groups: The three curves now intertwine seemingly at random and the systematic deviations disappear (figure~\ref{fig:folds}a; inset).

A better view of the systematic deviations is provided by the statistics of the location of the main peak, where the differences are most pronounced.  In figure~\ref{fig:folds}b we show histograms for the locations of the main peak
in the three  groups (same color coding as before), along with best gaussian fits to the histograms for the two extreme groups (all-$\alpha$ and all-$\beta$).  Despite the large overlap the shifts are quite apparent: The peak for all-$\alpha$ proteins is at about $85\,{\rm cm}^{-1}$, while that of beta-proteins is at 70 -- 75$\,{\rm cm}^{-1}$.  The fact that $\alpha$-helices are more compactly structured than $\beta$-sheets may account for the stiffer (higher frequency) peak for all-$\alpha$ proteins.

%
In the high-frequency range (and using all dofs), differences between $\alpha$- and $\beta$-rich proteins are observed near the range of amide vibration frequencies that have been extensively used to distinguish between $\alpha$-helix and $\beta$-sheet in protein infrared (IR) spectroscopy~\cite{ Nevskaya1976, Krimm1986,Susi1986, Goormaghtigh1990,  Fen-NiFu1994, Cai1999, Cai2004,Yang2015}. 
Figure~\ref{fig:amide}(A) shows the vibrational spectra in Cartesian dofs of all-$\alpha$ proteins in a red curve, all-$\beta$ proteins in a blue curve, $\alpha/\beta$-proteins in a gray curve, and three amide regions (I, II, and III) in light gray bands.
Figure~\ref{fig:amide}(B) shows an enlarged view of the three amide regions with additional arrows that point at peaks in the frequency curves. 
Though the overall vibrational spectrum is universal for all globular proteins,
Figure~\ref{fig:amide} shows that there are small, yet noticeable local differences in the three amide regions
between all-$\alpha$ and all-$\beta$ proteins.
Remarkably, in each amide region, not only the general locations of the peaks   but also the magnitudes of the shifts between all-$\alpha$ and all-$\beta$ proteins as predicted by our method match well with results from infrared spectroscopy~\cite{Nevskaya1976, Krimm1986,Susi1986, Goormaghtigh1990,  Fen-NiFu1994, Cai1999, Cai2004,Yang2015}.

\subsection{Using the Vibrational Spectrum to Assess and Improve Theoretical Approaches}  
\label{hypref:subsec:improvetheory}

While $g(\omega)$ is universal for a given atomic potential function, one would expect to see different curves for different formulations and parameterizations of the potential.   There is, however, only one reality and the ``true'' shape of $g(\omega)$ can only be  decided by experiment.  We show below how this effect can help one choose
between different potential formulations.  The vibrational spectrum is also quite sensitive to different levels
of approximation (simplified models/potentials, restricted dofs) and this  can be exploited to assess the accuracy
of various simplified models 
and help fine-tune their parameters.

\paragraph{Sensitivity to different empirical potentials.}
\label{subsec:potentials}
\label{hypref:subsec:potentials}

In order to demonstrate the sensitivity of $g(\omega)$ to the potential function one uses for the analysis, we compute the vibrational spectrum in two different ways: (a)~With the atomic detailed CHARMM22 potential~\cite{MacKerell98} (Figure~\ref{fig:potential}, in solid line), and (b)~with the same potential but where the Van der Waals radii of the various atoms is {\em replaced} with the values from the L79 potential function~\cite{L79,Levitt85},
and with a uniform torsional spring constant, $K_{\phi} = 0.1\,{\rm Kcal}/{\rm mol}/{\rm rad}^2$ (in dashed line).  The shift of the main peak to smaller frequencies, in the latter case, and the overall shape of the curve is quite in agreement with the spectra of proteins obtained with the L79 potential from the start~\cite{Tirion15}.  (For simplicity, we performed the analysis in the restricted set of torsional dofs, which suffices for our purposes.)  This demonstrates quite cleanly the influence of different parameter values in different (or the same) potential function(s).  It also lends further support to our claim that the location of the main peak has to do largely with non-bonded interactions (see Section~\ref{universality:subsec}).

Which potential function gives a better parameterization of the Van der Waals radii?  This, and similar issues,
can only be decided by experiment.  Experimental spectra are hard to obtain, since their extraction often requires 
various uncontrolled assumptions and approximations, and they remain a challenge.  One recent {experimental} study,
employing ultrafast OHD-RIKES
spectroscopy~\cite{Giraud03} in the low-frequency spectrum, 
finds  a main peak at about  $80\,{\rm cm}^{-1}$ and a broader, diffuse peak  at around $300\,{\rm cm}^{-1}$ (locations marked in the figure by vertical dotted lines).  This seems to favor the CHARM22 potential formulation over that of L79.  Agreement between theory and experiment remains an elusive goal, though: We hope that theoretically obtained $g(\omega)$'s would spur experimental interest in searching to confirm the various peaks, and conversely, that the various peaks observed
by experiments would help fine-tune the theoretical potential functions, thus bringing greater understanding.

\paragraph{Sensitivity to different levels of approximation.}
\label{approxes:subsec}
\label{hypref:approxes:subsec}

The spectrum distribution $g(\omega)$ is also sensitive to various levels of approximation that are used in simplified NMA models.  We have already seen one obvious example of this in Section~\ref{torsional:subsec}, in the different spectra one obtains with the restricted set of torsional dofs vs.~the full complement of Cartesian dofs.  We now examine the effect of other common simplifications.  For simplicity we use once again torsional dofs only, since they suffice to capture the differences.

In figure~\ref{fig:models} we plot $g(\omega)$ as obtained from NMA with the full atomic CHARMM22 potential (black line), along with decreasing levels of approximation: sbNMA (in blue), ssNMA (red), and ANM (gray).
Since NMA requires minimization of the CHARMM22 potential function we use the minimized structures also in all the approximate techniques, so as to obtain a fair comparison.  (The effect of the starting configuration is discussed separately, in the next section.)  The quality of the approximate approaches is clearly reflected by the general trend: the better the approximation the better the fit of its $g(\omega)$ to that of NMA.

The best approximation, sbNMA, also yields the best fit:  The main peak is reproduced very faithfully and significant differences show only in the secondary peak, which seems shifted somewhat towards higher frequencies.  This makes sense, in view of the fact that the torsional terms in the potential (responsible for the secondary peak) are more heavily approximated in this technique than the non-bonded interaction terms (that shape
the main peak): The spring constant for the torsional terms is fixed at a one value, regardless of the amount of rotation, as opposed to the non-bonded spring constants whose values are a function of the distance between the interacting atoms.
The near perfect reproduction of the first peak confirms that electrostatic interactions, which are neglected in sbNMA, indeed contribute significantly less to the normal mode motions than the van der Waals interactions~\cite{Na14b}.
The ssNMA, that uses a smaller set of parameters than sbNMA, results in further deterioration of $g(\omega)$.  
Finally, the coarsest approximation, ANM, with only one universal spring constant for all interactions, yields the worst fit to the $g(\omega)$ of the original NMA.

One can use the $g(\omega)$ to improve the various approximations.  As an example, the maximum of the secondary peak of sbNMA, at about $290\,{\rm cm}^{-1}$, could be shifted to the NMA's maximum at $270\,{\rm cm}^{-1}$ by softening the torsional spring constants by a factor of $\sqrt{290/270}\approx0.87$ (since $\omega$ is proportional to the square root of the spring constant, and the secondary peak is mostly determined by torsional energy terms).  Indeed, this ploy succeeds in effecting the desired shift ({results not shown}), though the quality of the main peak somewhat deteriorates.  One could in principle fine-tune the various sbNMA parameters to achieve an {\em optimal} fit to $g(\omega)$ of the original NMA.

  It is less clear how to improve the fit of the main peak with ssNMA.  Juxtaposing
the results of sbNMA and ssNMA, it becomes apparent that detailed spring constants for non-bonded interactions,
dependent on atom type and distance, as in sbNMA~\cite{Na14b} (and ATMAN~\cite{Tirion15}), are crucial to a faithful reproduction of the main peak.  Likewise, the torsional spring constants are essential to a successful reproduction of the secondary peak.  For the simplest approximation, ANM, varying the single available spring constant would only result in an overall scaling of the frequencies, or the $\omega$'s.  In other words, the whole curve would compress or dilate uniformly, as
the spring constant is softened or strengthened.  Thus, softening the spring constant might achieve a better
fit of the maximum location of the main peak, that would then shift to the left, however, this would also result
in further narrowing of the peak (which seems already too narrow in comparison to NMA).  In short, it seems rather impossible to achieve a satisfying fit with such a simple approximation.

In closing this section, we note that the quality of the various approximations has formerly been assessed by comparing
individual modes, rather than the distribution of their frequencies, as suggested here.  For example, modes comparison has been used by
Na and Song~\cite{Na15a} to conclude that a good quality approximation, such as sbNMA or ssNMA, requires geometric terms that maintain proper bond lengths and bond angles, 
 distance-dependent van der Waals based spring constants for non-bonded interactions, plus torsional spring constants, as a minimal set of parameters.   Interestingly, Tirion and ben-Avraham~\cite{Tirion15} reached the same conclusion in their development of the closely related ATMAN approach, but using  $g(\omega)$ as a guide.  Clearly the two techniques, comparing individual modes and comparing frequency distributions, have their own problems and merits and are complementary to one another, enriching our chest of theoretical tools.

\subsection{How Input Structures Affect the Vibrational Spectrum}
\label{subsec:input_structure}
\label{hypref:subsec:input_structure}

A great advantage of simplified potential functions of the type introduced by Tirion~\cite{Tirion96} in 1996, such as ANM, sbNMA, ATMAN, etc., is that they require no minimization: The potential function is at a minimum at the outset, regardless of
the protein's given configuration.  Thus, using such potentials one can obtain the normal modes of a protein
for any number of different {\em starting configurations}, or {\em input structures}.  It is well known that as
long as the input structures do not differ by a large amount, the first few slowest modes
remain quite unchanged (for a recent detailed, quantitative study, see Na and Song~\cite{Na15c}).  We here
show, however, that the overall distribution of mode frequencies, $g(\omega)$, {\em is} affected by different input
structures.  The question then arises of what is the {\em proper} input structure for a normal mode analysis.

To demonstrate the effect, in 
figure~\ref{fig:input} we show the spectra obtained with sbNMA with two different input structures: {(i)}~The PDB configurations of the proteins, and {(ii)}~the configurations obtained by minimizing the 
CHARMM22 potential energy.  Recall that the sbNMA parameters are determined from the CHARMM22 potential,
but being a Tirion-type potential it allows us to obtain spectra with the two different input sets.  (In contrast,  
a full detailed potential such as CHARMM22 must first be minimized and is limited to only the minimized structures.)  
For the minimized structures, sbNMA and NMA using CHARMM22 obtain very similar spectra, as we have already demonstrated in Section~\ref{approxes:subsec} (the curve for CHARMM22 is included in the figure, as a reminder).
The point of this plot is the significant differences between the spectra of {(i) and (ii)}: For example, the maximum of the main peak, located at about $80\,{\rm cm}^{-1}$ for the minimized structures, shifts to about $50\,{\rm cm}^{-1}$
for the (non-minimized) PDB inputs. 
{On one hand, universality as shown in figure~\ref{fig:universal} indicates the spectrum is not protein-specific, arising from structural properties common to globular proteins in general. On the other hand, results from figure~\ref{fig:input} imply that the spectrum depends quite strongly on whether the structures are minimized or not. 
(The spectra obtained for the same 135 structures but without energy minimization  are universal as well. See Figure S1 in Supplemental Information). 
The two are not contradictory to each other. 
Their difference, specifically the shift of the first peak, can be understood as follows. First, recall that the spectrum at the low frequency end is contributed mainly by non-bonded interactions, especially the van der Waals interactions, which are sensitive to inter-atomic distances. 
Second, energy minimization causes the structures to relax according to the force field. The structure change inevitably alters the inter-atomic distances and consequently the van der Waals terms in the potential function. The change in the latter in turn mostly determines the shift in the location of the first peak. At the high-frequency end, however, there is little difference between spectra of minimized structures and non-minimized structures (see Figure S2 in Supplemental Information). 
}

So, what is the proper starting configuration for a normal mode analysis?  Ideally, one would like to use the equilibrium configuration of a {\em single} protein in its natural state, but the PDB configurations are obtained from
{\em crystal} structures and it is not quite clear whether these two are the same.  Taking into account that the $g(\omega)$ from the minimized structures agree better with the experimental results of Giraud et al.,~\cite{Giraud03} (main peak at about $80\,{\rm cm}^{-1}$), there are two possibilities: If the CHARM22 potential is to be trusted then this suggests that the PDB crystal structure is different from a single protein's equilibrium structure; conversely,
if the PDB crystal structure is the same as the protein's equilibrium structure, this suggest that CHARMM22 
potential is not quite right and its various parameters need to be adjusted.  On the one hand, there is enough
reason to suspect that the crystal packing distorts equilibrium configurations.  For example, the configuration of G-actin,
as obtained from its crystal form, needs to be distorted significantly in order to fit with the structure derived in the different packing of F-actin filaments~\cite{Tirion95}.  On the other hand, experimental spectra are difficult to interpret and the results are far from uniform.  It might very well be that the PDB structures ought to be trusted
more than the CHARMM22 parameters.  If the latter is the case, normal modes analysis must proceed from
undistorted PDB structures (using a Tirion-type potential such as sbNMA, or ATMAN), as argued by Na and Song~\cite{Na15c}. 
The actual answer seems important, either way.


\section{Conclusion and Discussion}
\label{conclusion:sec}
\label{hypref:conclusion:sec}

In this work, 
we have shown that the density of modes in the vibrational spectrum of globular proteins is universal: 
The density of modes of different globular proteins, when properly normalized, tend to aggregate around one universal curve. 
We find this universality to be true not only for the low frequency range and for the restricted set of torsional dofs, as observed in earlier studies~\cite{Tirion93,dba93}, but for the whole frequency spectrum and for the full complement of dofs available to the proteins' atoms. 
This surprising result is highly significant, in that it implies that the universal patterns of the spectrum,  its turns and peaks, are not protein-specific but rather force-field specific, arising from structural  properties and inter-atomic interactions common to globular proteins in general.

The universality of the spectrum density and the fact that the actual $g(\omega)$ curve depends on the empirical potential used for the normal modes analysis calls for a serious two-way dialogue between theory and experiment: Experimental spectra {of proteins} could now guide the fine tuning of theoretical empirical potentials, and the various features and peaks observed in theoretical studies --- being universal, and hence now rising in importance --- would hopefully spur experimental confirmation.

The characterization of the typical {\em fluctuations} from the average $g(\omega)$ paves the way to the interpretation of {\em salient features} in the spectra of {\em individual} proteins, thus promising to fulfill a decades-old goal of continued work on normal mode analysis.  That this is possible, in principle, was clearly demonstrated by the discernible differences in the spectra of proteins of different fold families (see Section~\ref{subsec:folds}).

The universality of $g(\omega)$ also provides us with an exquisite tool for the assessment of various approximation approaches.  We have thus seen that in order to obtain a faithful resemblance of the  NMA spectrum, an approximate technique must include, at the very least, spring constants
for the non-bonded interactions that are atom-type dependent and distance-dependent, and include energy terms for changes of torsional and dihedral dofs, as done in sbNMA~\cite{Na14b}, or ATMAN~\cite{Tirion15}.  {This level of accuracy is indispensable
for highly sensitive tasks such as 
finding the different vibrations of closely related crystal isoforms of a protein~\cite{Tirion15b} and can improve, in general, the results of many normal mode-based studies, such as
identifying folding cores~\cite{Bahar98} or hot-spot residues~\cite{Ozbek13}. }
  In the opposite extreme, the simplest ANM approximation, while  useful for predicting the general shape of the  slowest modes, cannot simultaneously account for both the location and the width of even the main peak of $g(\omega)$.

What is the source of the universality we observe in $g(\omega)$?  For low frequency modes, it has been argued that the coherent motions of large domains of a protein involve mainly interactions between atoms of adjacent domain surfaces.  Those interactions average out in the same way, for all proteins, simply because the number of interacting pairs is large and one can then invoke the central limit theorem to describe their combined effect.
This argument does not work, however, for higher frequencies, where the coherence length of the modes is tiny and 
the moving components involve but a few atoms.  One possibility is that, due to the very different stiffnesses of torsional changes, angle bending, and bond stretching, the three elements dominate different parts of the spectrum:  torsional terms in the low frequency range of $0$ -- $500\,{\rm cm}^{-1}$, angle bending in the intermediate range of
$500$ -- $2000\,{\rm cm}^{-1}$, and bond stretching in the high-frequency range, above $2000\,{\rm cm}^{-1}$ (see
figure~\ref{fig:peaks}).  Different proteins may have similar percentages of the various angle and bond types, explaining the universality in the mid- and high-frequency range.  
Only future work could unravel the full causes for the universality of $g(\omega)$, and whether the distinct stiffness magnitude for the three types of interactions play a decisive role in it.

Among  the many other interesting open questions left by this study we mention the precise relation between Cartesian and torsional dofs.  The densities of the two spectra look very similar, but there is an excess of modes
with Cartesian dofs (see figure~\ref{fig:Cartesian-torsional}). Quantitatively speaking,
it seems like that there are about 30\% more modes in Cartesian space, at the low frequency end. Why?
A possible explanation is that some Cartesian modes represent a mix of torsional and non-torsional motions (such as bond bending motions), as indicted by figure~\ref{fig:peaks}, but still oscillate at low frequencies.  Whether
this is the case, as well as a detailed comparison of the slow modes themselves in Cartesian and torsional dofs, is left for future work.

Is the universality of $g(\omega)$ true only for globular proteins or does it encompass other types of proteins? 
It would be interesting if the same spectrum density, or features of it, resulted also for non-globular proteins.   How would various ligands affect the spectrum density? In a recent work by Wynne and co-workers~\cite{Turton14}, the spectrum of vibrations was found to be modified after ligand binding (and the range of affected modes was postulated to play an important role in the ligand binding process as well).  Since proteins of different fold families exhibit discernible variations in the spectrum density, it is plausible that other types of proteins, and ligands,  would also affect the spectrum.  Is protein size more relevant than suggested by our results?  In the present study, we were limited to small- to medium-size proteins, for the ease of {all-atom NMA computations}. 
We expect an inverse correlation between protein size and deviations from universality --- the smaller the protein, the larger the deviation --- but we were unable to see that in our data.  A larger study is needed to prove or disprove this notion.  Ultimately, a full characterization of the fluctuations is important in order to correctly identify outlier proteins.  It would be interesting to study a few outlier proteins whose spectra truly differs from the average, beyond the expected fluctuations.

The question of the proper starting conformation for a normal mode analysis remains a genuine puzzle (Section~\ref{subsec:input_structure}).  Given an experimentally determined structure, should we first minimize it before performing NMA?
We have seen that the location of the main peak shifts significantly, from $80\,{\rm cm}^{-1}$ for potential-minimized structures, to about $50\,{\rm cm}^{-1}$ for the original crystal structures given by the PDB files.  We have postulated that part of the effect is due to the conformational distortions undergone by the proteins in the crystal packing. 
{Future spectral studies based on nuclear magnetic resonance (NMR) structures that are not influenced by crystal packing} 
might shed light on this issue.  
Whatever the answer, the fact that the input configuration makes a big difference in the outcome emphasizes the need for further development of good Tirion-type potential energy formulations, such as sbNMA and ATMAN, that are minimized at the outset.

An issue that we have left untouched in this study is the question of the precise nature of the spectrum of vibrations in the low-frequency range. In early work~\cite{Elber86,dba93} it was suggested that the low-frequency spectrum has an anomalous {\em spectral} dimension of $d_s\leq2$ (instead of 3, as expected for a three-dimensional crystal): This implies that the low-frequency spectrum behaves in power-law fashion, $g(\omega)\sim\omega^{d_s-1}$.  Later studies~\cite{Burioni04,Reuveni08} found a weak dependence of the spectral (and fractal) dimension with protein size.  The power-law (and anomalous dimensions) interpretation has been contested by Etchegoin and N\" ollmann~\cite{Etchegoin98,Nollmann99}, who maintained that the low-frequency spectrum rather  fits a log-normal distribution and is better explained by the analogous behavior in glasses (their analysis, however, relied on spectra obtained with only torsional dofs).  We have not attempted to delve into this argument, mostly due to the limited range of sizes of our proteins.  A future study, involving a larger dataset and heavier proteins, and using the NMA method with an all-atom potential and the full complement of dofs, as done in the present work, will shed much needed light on this interesting problem.


\ack

We thank Dr.~Monique M.~Tirion for many useful discussions and for a critical reading of the manuscript.
Funding from National Science Foundation (CAREER award, CCF-0953517) is gratefully acknowledged.


\clearpage
\bibliographystyle{unsrt}
\bibliography{reference}


\section*{}
\pagebreak
\section*{Figure Legends}

\subsection*{Figure 1.}
({\bf A}) The size distribution of the 135 proteins used in this work; {({\bf B}) the extent of structure deviations caused by energy minimization among the same 135 proteins.}

\subsection*{Figure 2.}
{Universality of the density of vibrational modes of globular proteins.}
The black line shows the  average of the 135 proteins in the dataset.
The fluctuations from the average are represented by the color bands, demarcating the fraction of proteins
that are included within various ranges: 25--75 percentile (orange), 5--95 percentile (red), 0--100 percentile (gray).
The bin size $\Delta\omega$ used here for computing the density of modes is $10\,{\rm cm}^{-1}$.

\subsection*{Figure 3.}
Spectrum of vibrations for Cartesian vs. torsional dofs for four example proteins: 3NBC ({\bf A}), 3RHB ({\bf B}), 2QCP ({\bf C}), and 3MP9 ({\bf D}).  The averaged density of modes (over all the proteins in the dataset) for Cartesian and torsional dofs are represented respectively by the dashed and solid red curves.

\subsection*{Figure 4.}
{Relative contribution of the various interaction terms to the vibrational spectrum.}
Inset: The main and secondary peaks shown in more details.

\subsection*{Figure 5.}
{The torsional-dofs spectrum of vibrations with and without various interaction terms.}
The black line is the spectrum of NMA while the gray line that of sbNMA, a model that closely resembles NMA. When the torsional interaction term is removed from sbNMA's potential, the second peak disappears. The first peak disappears (or reshapes significantly) when non-bonded interaction term is removed.

\subsection*{Figure 6.}
{Vibrational spectra and statistics of the main peak location for different protein folds.}
({\bf A})~Vibrational spectra for different protein folds: 42 all-$\alpha$ proteins (red), 37 all-$\beta$ proteins (blue), and 56 $\alpha/\beta$-proteins (gray). Notice the systematic shifts in the main peak and the far slope of the secondary peaks.
Inset: Plots of $g(\omega)$ when equivalent number of proteins are assigned {\em randomly} to the three groups as in the main plot. The systematic deviations disappear.  
({\bf B})~Statistics of the location of the main peak for the three groups of protein (color coding same as ({\bf A})) roughly fits a Gaussian distribution (curves shown for all-$\alpha$ and all-$\beta$ proteins only) and clearly demonstrate the differences between the various types.

\subsection*{Figure 7.}
{Vibrational spectra of amide groups for different protein folds.}
({\bf A}) shows the vibrational spectra for different protein folds in the range of amide vibration frequencies: 42 all-$\alpha$ proteins (red), 37 all-$\beta$ proteins (blue), and 56 $\alpha/\beta$-proteins (gray). The frequency range of amide I, II, and III are highlighted in light gray bands.
({\bf B}) zooms in on the three amide regions. Arrows point out peaks of frequency curves of all-$\alpha$ and all-$\beta$ proteins in the amide regions.

\subsection*{Figure 8.}
{Vibrational spectra obtained with the CHARMM22 potential (solid) and the approximated L79 potential (dashed).}
The two vertical dotted lines mark the locations of the first and second peaks observed experimentally~\cite{Giraud03}, 
which favor the CHARMM22 potential formulation over that of L79.

\subsection*{Figure 9.}
{The vibrational spectra obtained by the original NMA and various simplified models.}
Vibrational spectra provide a critical assessment of the quality of the simplified models.

\subsection*{Figure 10.}
{Dependence of the vibrational spectrum on input structures.}
{The spectrum of sbNMA varies significantly (for example, in the location of the main peak) when different input structures are used. The spectrum of NMA is shown in the background as a reference. }

\pagebreak

\section*{}
\begin{figure}
\centerline{\includegraphics[width=\textwidth]{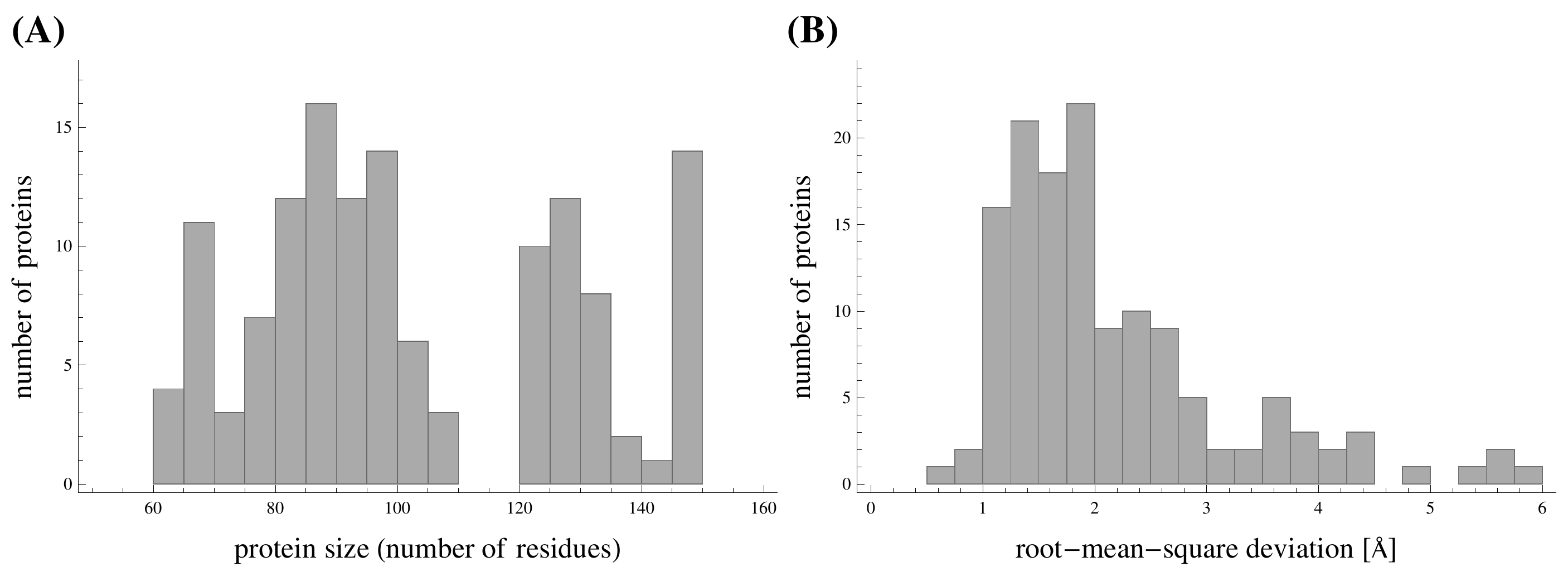}}
\caption{}
\label{fig:sizes}
\end{figure}

\section*{}
\begin{figure}
\centerline{\includegraphics[width=\textwidth]{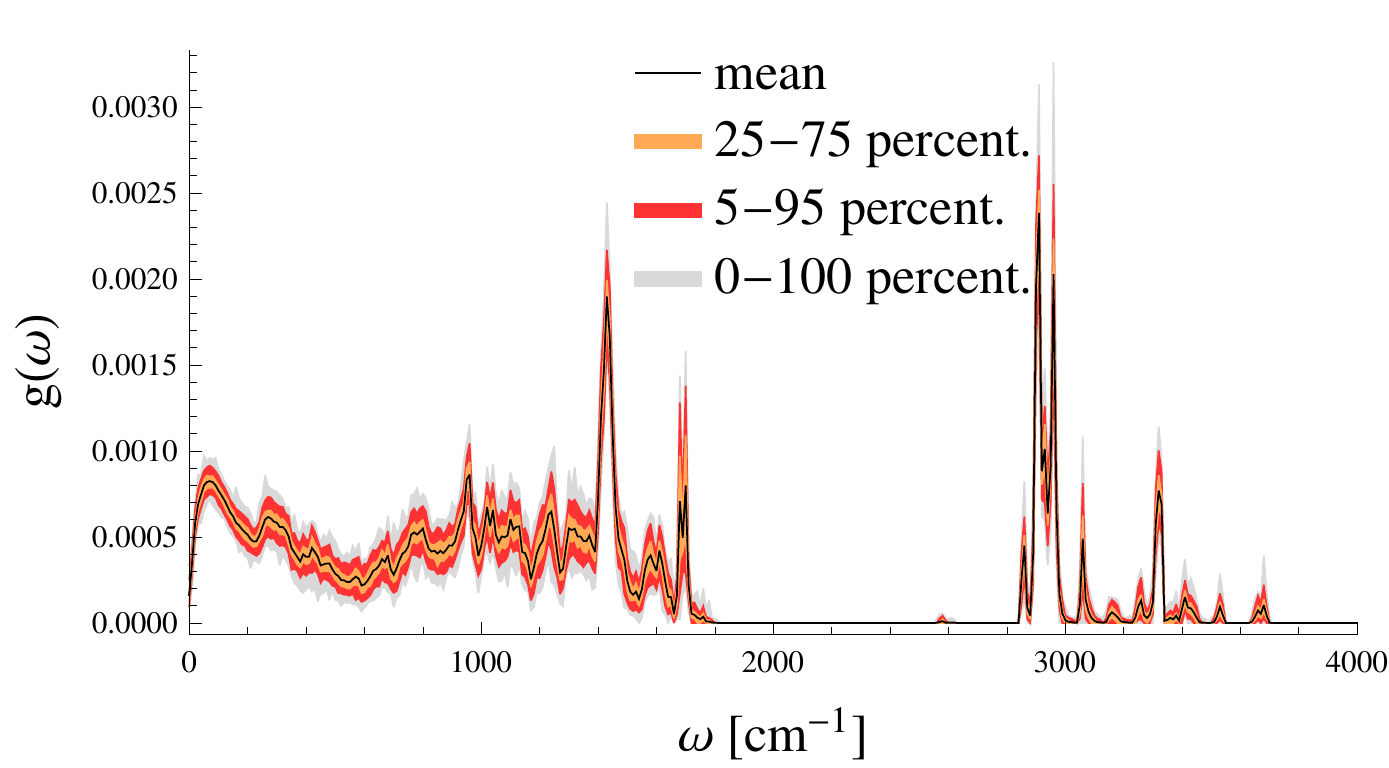}}
\caption{}
\label{fig:universal}
\end{figure}

\begin{figure}
\centerline{\includegraphics[width=\textwidth]{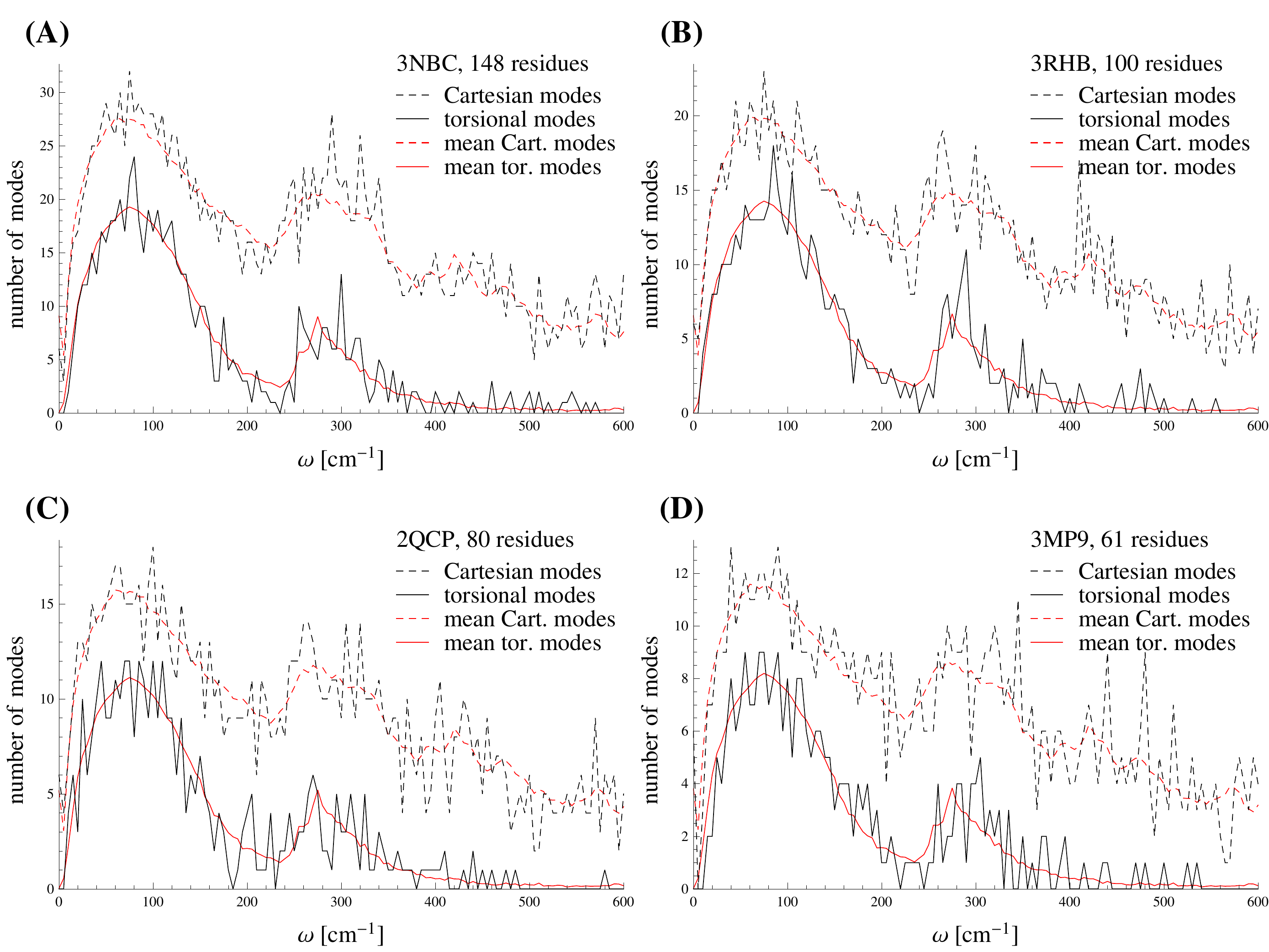}}
\caption{}
\label{fig:Cartesian-torsional}
\end{figure}

\begin{figure}
\centerline{\includegraphics[width=\textwidth]{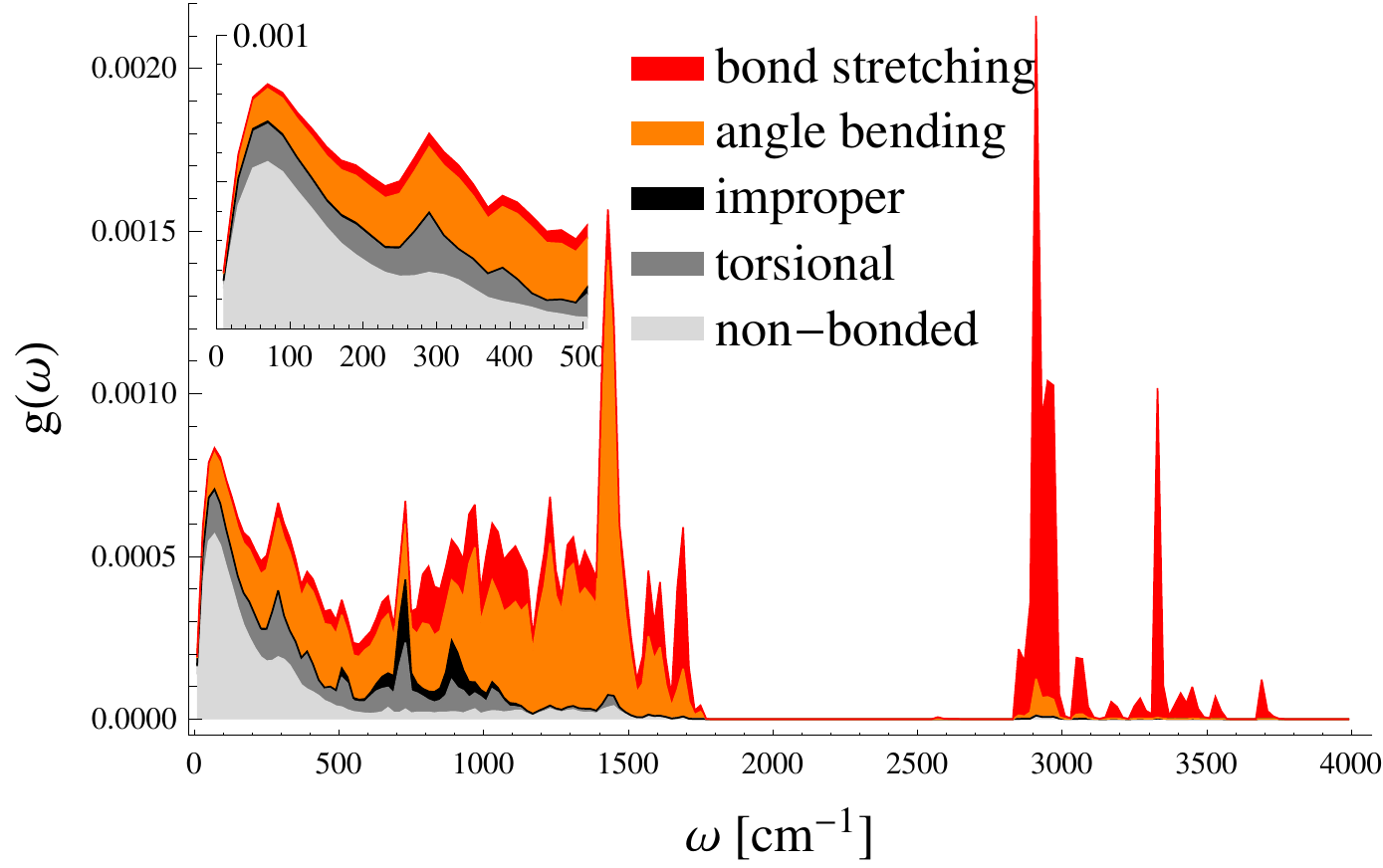}}
\caption{}
\label{fig:peaks}
\end{figure}

\begin{figure}
\centerline{\includegraphics[width=\textwidth]{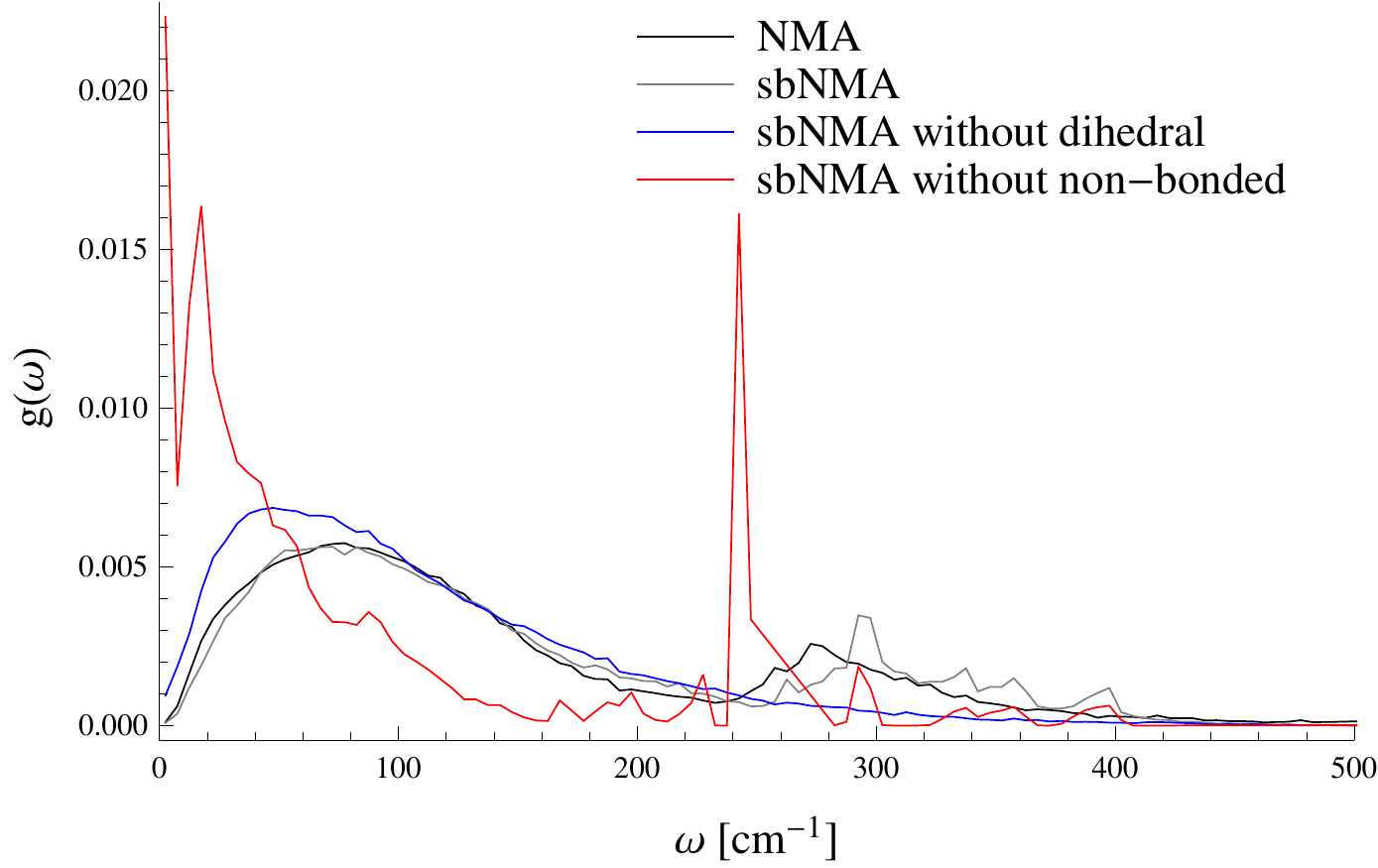}}
\caption{}
\label{fig:torsional}
\end{figure}

\begin{figure}
\centerline{\includegraphics[width=\textwidth]{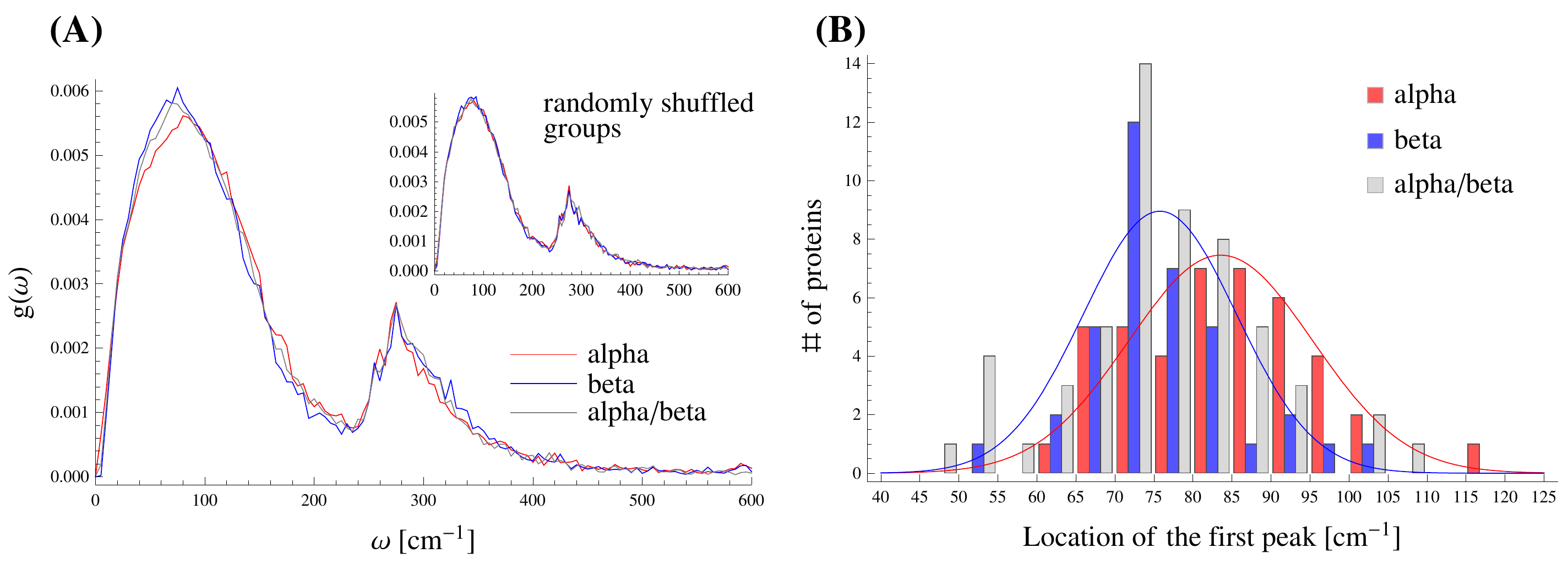}}
\caption{}
\label{fig:folds}
\end{figure}

\begin{figure}
\centerline{\includegraphics[width=\textwidth]{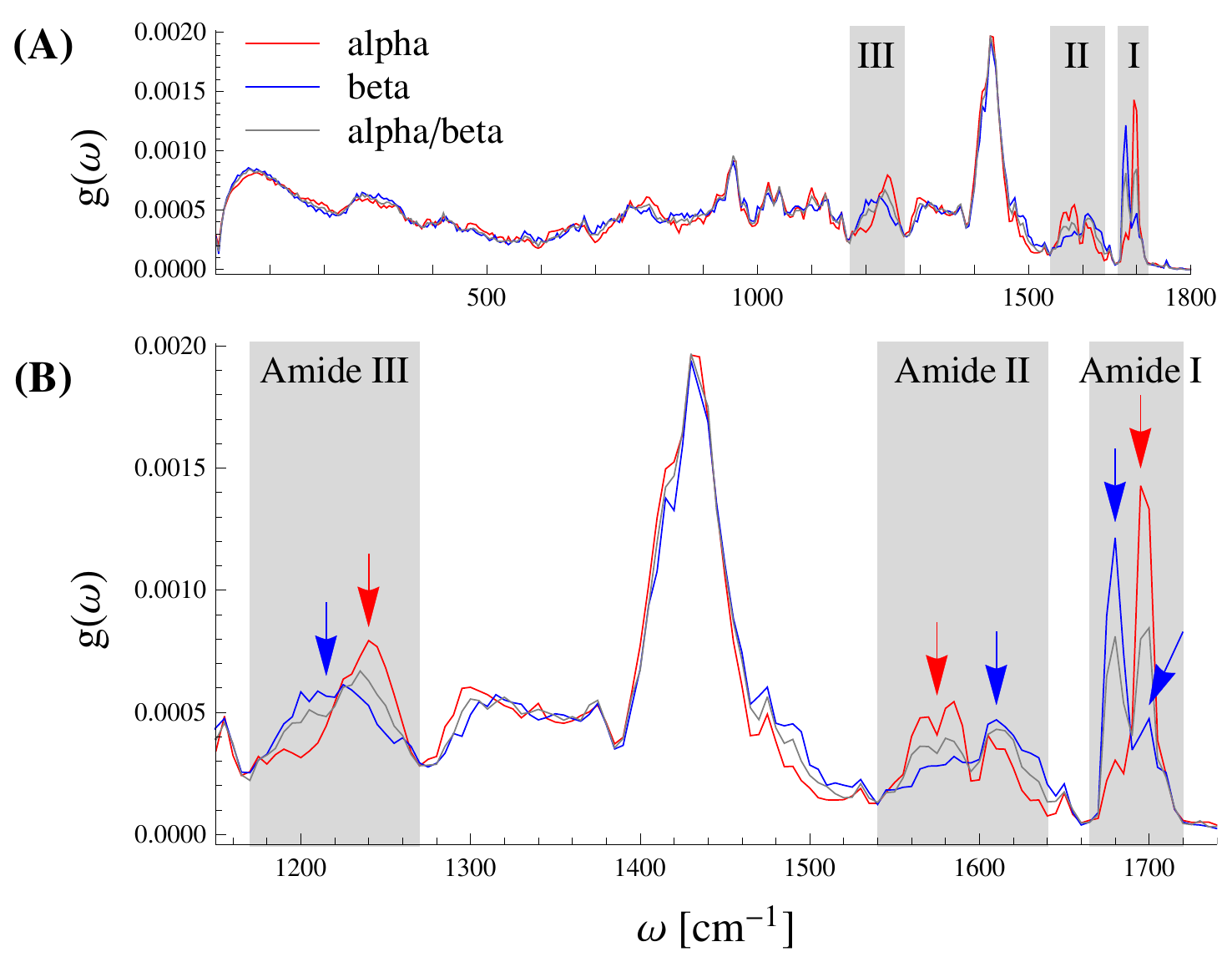}}
\caption{}
\label{fig:amide}
\end{figure}

\begin{figure}
\centerline{\includegraphics[width=\textwidth]{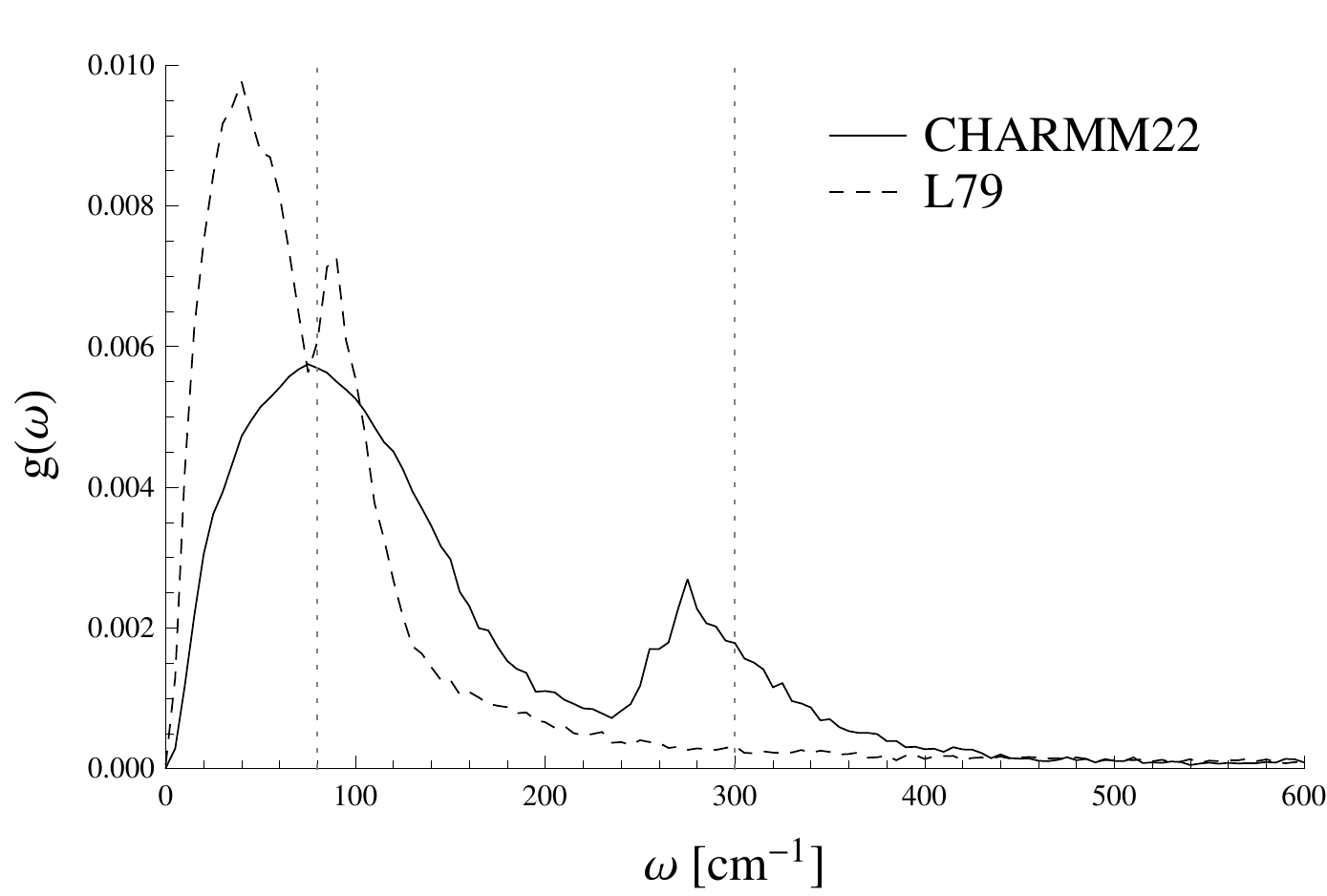}}
\caption{}
\label{fig:potential}
\end{figure}

\begin{figure}
\centerline{\includegraphics[width=\textwidth]{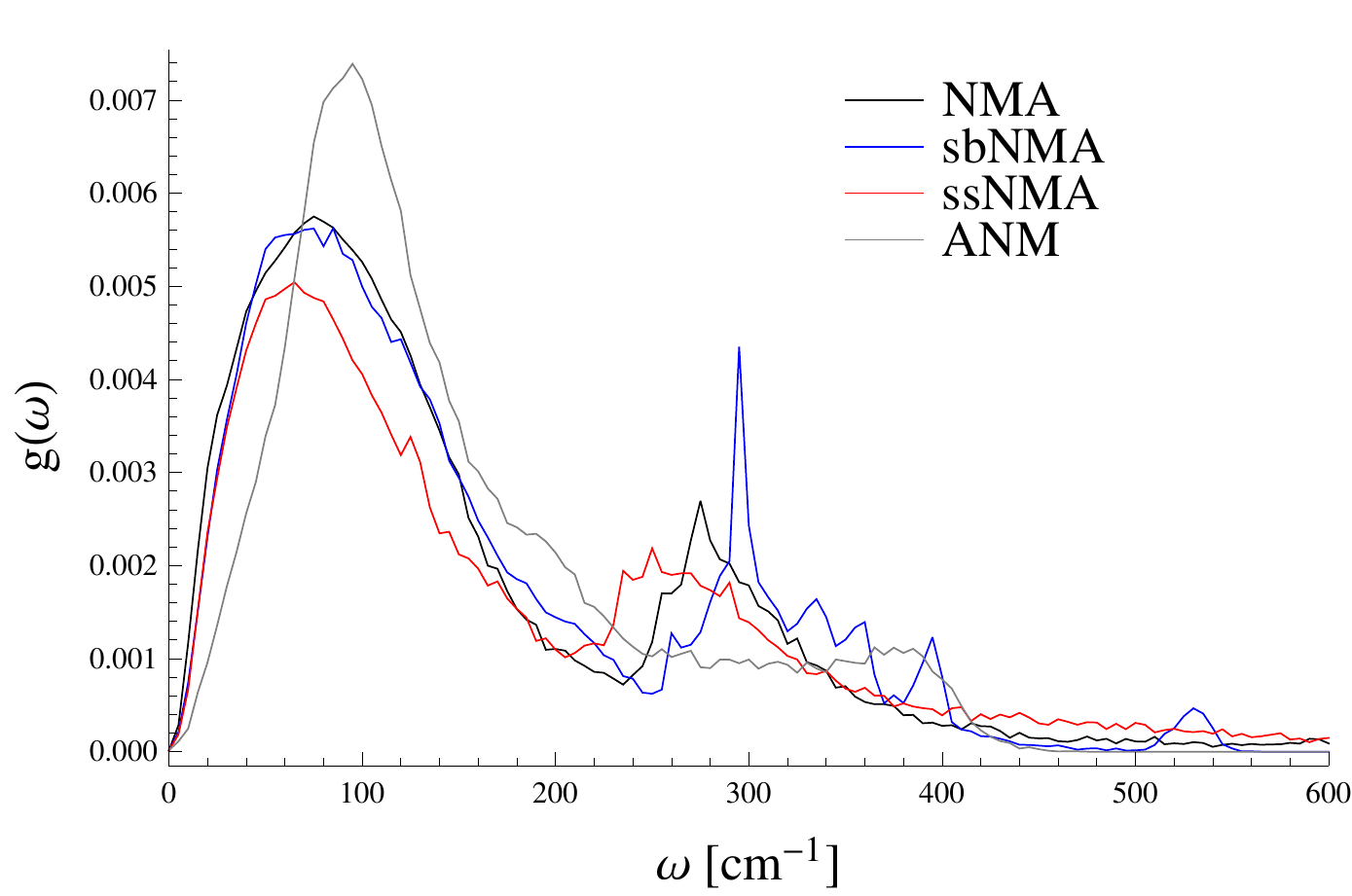}}
\caption{}
\label{fig:models}
\end{figure}

\begin{figure}
\centerline{\includegraphics[width=\textwidth]{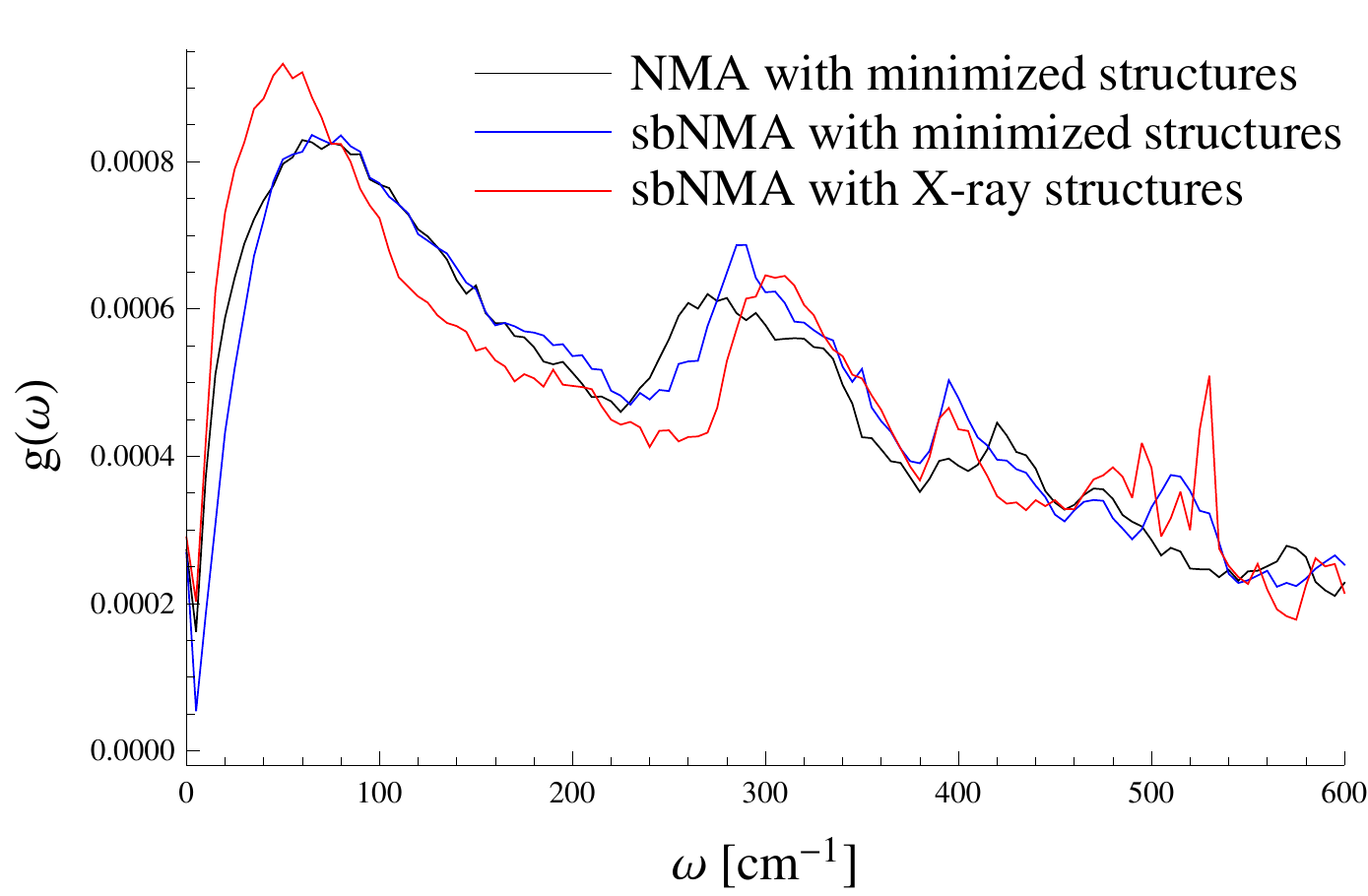}}
\caption{}
\label{fig:input}
\end{figure}

\end{document}